\documentclass[aps,pra,twocolumn,amsfonts,amssymb,amsmath,showpacs,
floatfix,nofootinbib,groupedaddress,superscriptaddress,]{revtex4}

\usepackage{mathrsfs}
\usepackage{amsfonts}
\usepackage{amstext}
\usepackage{amsmath}
\usepackage{amssymb}
\usepackage{bm} \usepackage[utf8]{inputenc}
\usepackage[T1]{fontenc}
\usepackage{CJK}
\usepackage{bm}
\usepackage{bbm}
\usepackage[dvips]{graphicx}
\def\qed{\leavevmode\unskip\penalty9999 \hbox{}\nobreak\hfill
     \quad\hbox{\leavevmode  \hbox to.77778em{%
              \hfil\vrule   \vbox to.675em%
               {\hrule width.6em\vfil\hrule}\vrule\hfil}}
     \par\vskip3pt}

\begin{document}
\title{Conversion of Gaussian states under incoherent Gaussian operations}
\author{Shuanping Du}
\email{shuanpingdu@yahoo.com} \affiliation{School of Mathematical
Sciences, Xiamen University, Xiamen, Fujian, 361000, China}
\author{Zhaofang Bai}\thanks{Corresponding author}
\email{baizhaofang@xmu.edu.cn} \affiliation{School of Mathematical
Sciences, Xiamen University, Xiamen, Fujian, 361000, China}

\begin{abstract}
The coherence resource theory needs to study the operational value and efficiency which can be broadly formulated as the question: when can one coherent state be converted into another under incoherent operations.
We answer this question completely for one-mode continuous-variable systems by characterizing conversion of coherent Gaussian  states under incoherent Gaussian operations in terms of their first and second moments. The no-go theorem of purification of coherent Gaussian states is also built. The structure of incoherent Gaussian operations of two-mode continuous-variable systems
is discussed further and is applied to coherent conversion for pure Gaussian states with standard second moments.
The standard second moments are images of all second moments under local linear unitary Bogoliubov operations.
As concrete applications, we obtain some peculiarities of a Gaussian system: (1) There does not exist a maximally coherent Gaussian state which can generate all coherent Gaussian states; (2) The conversion between pure Gaussian states is reversible; (3) The coherence of input pure state and the coherence of output pure state are equal.

\end{abstract}


\maketitle


{\it \bf I. Introduction}

\vspace{0.1in}Manipulating physical systems always suffers from practical restrictions which limit the control we can exert. It is, e.g., extremely
difficult to exchange quantum systems undisturbed over long distances \cite{3Hor}.
In order to manipulate spatially separate subsystems effectively within the resource theoretic framework, this restricts us to local operation and classical communication (LOCC).  Under these operations, we have to prepare a certain kind of states, i.e., separable states. The states which can not be produced by LOCC are entangled. The entanglement is the key resource that allows to implement operations such as quantum state teleportation to obtain perfect quantum state conversion by consuming entanglement \cite{Benett}. The restrictions are vital in quantum communication and quantum technology and also drive a deep understanding of the fundamental laws of nature
\cite{3Hor,Ein, VPRK, Virman}.

As entanglement of pure states is among the manifestations
of the superposition principle, one can naturally see the phenomenon of coherent superposition as a valuable resource. Recently, the resource theory of quantum coherence
has attracted much attention, and various efforts are made to build the coherence resource theory \cite{Abe,Bau,Win,Strelt,Ben,Lud,Chi}. In this frame, free operations corresponding to LOCC in entanglement theory are incoherent operations $(\text{IO}s)$ that  can be interpreted as a measurement which can not create coherence even if one applies postselection on the measurement outcomes \cite{Bau}.

One of the central question in coherence resource theory is conversion of coherent states . It is aimed to study whether $\text{IO}s$ can introduce an order on the set of coherent states, i.e., whether, given two coherent states $\rho$ and $\sigma$,
either $\rho$ can be transformed into $\sigma$ or vice versa. The answer to this question determines the value of coherent states in technological applications.
The question has been solved for pure state case \cite{DBG,GT1,GT2} and for qubit state case \cite{Gourx, Benx, HLS}. More recently, the conversion between pure states and mixed states is characterized in \cite{Du2x, LZhou1x, FL1x}.

All the above results for  conversion of coherence states
are implicitly assuming a discrete-variable system (finite dimensional system).
Note that the first framework for understanding quantum coherence is quantum
optics which must require quantum states in a continuous-variable system (infinite dimensional system), especially the Gaussian states which have arisen to a privileged
position  in  continuous-variable quantum information \cite{Glaux, Sudx, Braux, Weedx}.
The primary tool for analyzing Gaussian states is Gaussian operations.
  Indeed, Gaussian operations correspond exactly to those operations
that can be implemented by means of optical elements such
as beam splitters, phase shifts and squeezers together with homodyne measurements \cite{Gie,Eis3,Fiu}. Such operations are in
principle experimentally accessible with present technology
\cite{Leo}. Especially, Gaussian unitary operations can be realized as a
 passive operation, a  single-mode squeezing operation on each of the $n$ modes, and a  subsequent second
passive operation \cite{Eis2}. In fact, phase rotation, the simplest and most common Gaussian unitary operation is an optical implementation
which preserves coherence in the process of conversion of coherent states \cite{Weedx}. For the process of evolution of optical cat states,
coherence is consumed \cite{Glancy,Mzhang}.

In the outlook of \cite{Bau}, T. Baumgratz, etc. point out that coherence theory of Gaussian systems is needed. Closely mirroring the
development of entanglement theory, mathematical problems concerning continuity that are inevitably emerging
can be addressed by requiring energy constraints \cite{Eis1} or
by considering special, experimentally relevant, subclasses
such as Gaussian states \cite{Eis2}.

The study of coherence theory of Gaussian systems is moving ahead since the question is proposed \cite{Bau}.
Recently, coherence theory of Gaussian systems including  incoherent Gaussian states, incoherent Gaussian operations and coherence measures of Gaussian states is introduced \cite{Xu1,Danx,Xu2}. The main contribution of our paper
is  to discuss  conversion of coherence states in continuous-variable systems under incoherent Gaussian operations ($\text{IGO}s$).

The paper is organized as follows. In section II, we review some background material and
establish notations. In particular, we review the definitions and characterizations of incoherent Gaussian states and incoherent Gaussian operations obtained in \cite{Xu1}. In section III,  an explicit description of conversion of Gaussian states of one-mode continuous-variable systems is provided
and the no-go theorem of purification for Gaussian states is built.
In section IV, we furtherly study the structure of IGOs in two-mode continuous-variable systems and characterize conversion of one kind of important Gaussian pure states under IGOs. The last section is a summary of our findings.

\vspace{0.1in}

{\it \bf II. Background and notation }\vspace{0.1in}

Let ${\mathcal H}$ be
an infinite dimensional Hilbert space with fixed orthonormal basis $\{|n\rangle\}_{n=0}^{+\infty}$.
When we consider the $m$-mode continuous-variable systems ${\mathcal H}^{\otimes m}$, we adopt $(\{|n\rangle\}_{n=0}^{+\infty})^{\otimes m}$
as its reference basis.
For a quantum state $\rho$ on ${\mathcal H}^{\otimes m}$, the characteristic function of $\rho$ is defined as
$$\begin{array}{lll}
{\mathcal X}_\rho(\lambda) & = & tr(\rho D(\lambda)),\\
D(\lambda) & = & \otimes _{i=1}^m D(\lambda_i),\\
D(\lambda_i) & = & e^{(\lambda_i\widehat{a_i}^\dag-{\overline \lambda_i}\widehat{a_i})},
\end{array}$$
here $\widehat{a_i}$ and $\widehat{a_i}^\dag$ are the annihilation and creation operator in mode $i$, \quad $\lambda=(\lambda_1, \cdots, \lambda_m)^t, \quad  {\overline \lambda_i}$ denotes the complex conjugate of $\lambda_i$. A one-mode quantum state $\rho$ is called Gaussian state if its characteristic function
$${\mathcal X}_\rho(\lambda)=exp^{-\frac{1}{4}(\lambda_x, \lambda_y)\Omega V\Omega^t(\lambda_x, \lambda_y)^t-i(\Omega d)^t(\lambda_x, \lambda_y)^t},$$ where $\lambda_x={\mbox Re}(\lambda)$ and
$\lambda_y={\mbox Im}(\lambda)$ are the real and imaginary parts of $\lambda$, $\Omega=\left(\begin{array}{cc}
                                                                         0 & 1\\
                                                                         -1 & 0\end{array}\right)$, $d=(d_1, d_2)^t\in{\mathbb R}^2$, $V=\left(\begin{array}{cc}
                                                                         v_{11} & v_{12}\\
                                                                         v_{12} & v_{22}\end{array}\right)\in {\mathcal M}_2({\mathbb R})$.
Recall that $V$ is a  positive definite matrix satisfying $V+i\Omega\geq 0$ (all eigenvalues are nonnegative) and $\det V\geq1$, $\det V=1$ if and only if $\rho$ is pure \cite{Weedx}.  $d$ and $V$ are called the first and second momemt of $\rho$ which can  describe Gaussian state $\rho$ completely. So $\rho$ can be usually written in $\rho(V,d)$.

The quantification of coherence is fundamental in the resource theory of quantum coherence.
For a given coherent Gaussian state, it is important to ask the amount of
coherence it has. Inspired by the idea of discrete-variable systems \cite{Bau}, researchers have built the framework for quantifying coherence of Gaussian states \cite{Xu1,Danx,Xu2}. For the convenience of reader, we give a brief overview of results in \cite{Xu1}.
The incoherent Gaussian states are defined as diagonal Gaussian states.   The set of incoherent Gaussian states will be
labelled by ${\mathcal I}$. The nondiagonal Gaussian states are called coherent Gaussian states. In fact, ${\mathcal I}$ consists of all  thermal states \cite{Xu1}.
A thermal state has the form $$\rho_{th}({\overline n})=\sum_{n=0}^{+\infty} \frac{{\overline n}^n}{({\overline n}+1)^{n+1}}|n\rangle\langle n|,$$ ${\overline n}=tr({\widehat a}^\dag\widehat{a}\rho_{th}(\overline n))$ is the mean photon number. A Gaussian operation is incoherent if it maps  incoherent Gaussian states into incoherent
Gaussian states.
A one-mode
incoherent Gaussian operation is fully described by $$(T,\  N), \  T=tO,\  N=\left(\begin{array}{cc}
                                                                                               \omega & 0\\
                                                                                                 0 & \omega \end{array}\right),$$
where $t$ is a real number, $O$ is a $2\times 2$ real orthogonal matrix $(OO^t=O^tO=I)$ and $\omega\geq|t^2\det O-1|$ \cite{Xu1}. For a Gaussian state $\rho(V,d)$, it performs on $\rho(V,d)$ and obtain a
Gaussian state with the first and second moment as follows:  $$ d\mapsto Td, \hspace{0.1in} V\mapsto TVT^t+N.$$

For  $m$-mode case,  every Gaussian state $\rho(V,d)$ is described by its first and second moment $d$ and $V$, where $d$ is a $2m$ dimensional column vector,
$V$ is a $2m\times 2m$ real positive definite matrix with $V+i\Omega\geq 0$,  $\Omega=\oplus_m \left(\begin{array}{cc}
                                                                         0 & 1\\
                                                                         -1 & 0\end{array}\right)$ \cite{Weedx}.
Furthermore $\det V\geq 1$ and $\det V=1$ if and only if $\rho(V,d) \quad\text{is pure}$. In \cite{Xu1, Xu2}, Xu has given detailed characterizations of incoherent Gaussian states and incoherent Gaussian operations (IGOs).
The incoherent Gaussian states have the form $\otimes_{j=1}^m \rho_{th}(v_j), v_j$ is the symplectic eigenvalue of $\rho_{th}(v_j)$.
A Gaussian operation $\Phi(T, N, d)$ is incoherent if and only if $$\left.\begin{array}{lll}
                                                                     d & = & 0,\\
                                                                      T & =& \{t_jO_j\}_{j=1}^m\in {\mathcal T}_{2m},\\
                                                                      N & = & \oplus_{j=1}^m \omega_j I_2,\\
                                                                      \omega_j &\geq &|1-\sum_{k, r(k)=j} t_k^2\det O_k|,\forall j,\end{array}\right.$$
where $t_j, \omega_j\in {\mathbb R}$, $O_j$ is a $2\times 2$ real orthgonal matrix $(O_jO_j^t=I_2)$, ${\mathcal T}_{2m}$ denotes the set of $2m\times 2m$ real matrices such that, for any $T\in {\mathcal T}_{2m}$, the $(2j-1,2j)$ two columns have just one $2\times 2$ real matrix $t_jO_j$ located in $(2r(j)-1, 2r(j))$ rows for $\forall j$, $r(j)\in
\{k\}_{k=1}^m$, and other elements are all zero. For a Gaussian state $\rho(V,d)$, it performs on $\rho(V,d)$ and obtain
a Gaussian state with the first and second moment as follows:  $$ d\mapsto Td, \hspace{0.1in} V\mapsto TVT^t+N.$$

Based on the definition of incoherent Gaussian states and incoherent Gaussian operations ($\text{IGO}s$),
any proper coherence measure $C$ is a non-negative function and
must satisfy the following
conditions:

$(C1)$ $C(\rho)=0$ for all $\rho\in{\mathcal I}$;

$(C2)$ Monotonicity under all incoherent Gaussian operations ($\text{IGO}s$) $\Phi$:
$C(\Phi(\rho))\leq C(\rho)$,

$(C3)$ Non-increasing under mixing of Gaussian states:
$C(\sum_jp_j\rho_j)\leq \sum_jp_jC(\rho_j)$ for any set of Gaussian states
$\{\rho_j\}$ and any $p_j\geq 0$ with $\sum_jp_j=1$.

Furthermore, the relative entropy measure has been provided by $$C_R(\rho)=\inf _{\delta\in {\mathcal I}} S(\rho||\delta),$$ $S(\rho||\delta)=tr(\rho\log_2\rho)-tr(\rho\log_2\delta)$ is the relative entropy.

For Gaussian states $\rho(V_1,d_1),\sigma(V_2,d_2)$, if there exists an incoherent Gaussian operation $\Phi$ such that $\Phi(\rho)=\sigma$, we denote it by  $\rho(V_1,d_1)\xrightarrow {\text IGO}\sigma(V_2,d_2)$. And  call $\rho$ and $\Phi(\rho)$ to be the input state and output state, respectively.

\vspace{0.1in}

{\it \bf III.  Conversion of  Gaussian states of one-mode continuous-variable systems}

Our first result provides a complete classification of conversion for pure Gaussian states.
This offers an  affirmative answer to the open question on coherence conversion in
continuous-variable systems \cite{Bau,DBG}. This question is to study how
can we determine if there exists an $\text{IGO}$ $\Phi$ such that $\Phi(\rho(V_1,d_1))=\sigma(V_2,d_2)$ for
pure Gaussian states $\rho(V_1,d_1)$ and $\sigma(V_2,d_2)$.

\vspace{0.1in}{\bf Theorem 3.1.} {\it For pure Gaussian states $\rho(V_1,d_1),\quad \sigma(V_2,d_2)$, $\rho(V_1,d_1)\xrightarrow {\text IGO} \sigma(V_2,d_2)$  if and only if there exists an phase rotation operator
$R(\theta)=\left(\begin{array}{cc}
                \cos\theta & \sin\theta\\
                -\sin\theta & \cos\theta\end{array}\right)$ for some $\theta\in{\mathbb R}$ such that $V_2=R(\theta) V_1R(\theta)^t, d_2=R(\theta)d_1$ or $V_2=I$ and $d_2=0$ ($\sigma(V_2, d_2)=|0\rangle\in {\mathcal I})$.}
\vspace{0.1in}

By Theorem 3.1, we obtain the following key peculiarities of Gaussian continuous-variable systems.

(i) $ \rho(V_1,d_1)\xrightarrow {\text IGO}\sigma(V_2,d_2)\text{ iff } \sigma(V_2,d_2)\xrightarrow {\text IGO} \rho(V_1,d_1)$ for coherent pure Gaussian states $\rho(V_1,d_1), \sigma(V_2,d_2)$.

(ii) For any coherence measure $C$, $C(\rho(V_1,d_1))=C(\sigma(V_2,d_2))$ if  $ \rho(V_1,d_1)\xrightarrow {\text IGO}\sigma(V_2,d_2)$. This implies that the coherence of input state is equal to the coherence of output state.
It shows frozen behavior of coherence in Gaussian dynamical systems. Frozen coherence in discrete-variable systems studied in \cite{Adesoxx} is to discuss when $C(\rho)=C(\Phi(\rho))$ holds true. Frozen coherence in Gaussian dynamical systems  is also listed as an open question in summary of \cite{Xu1}.

(iii) There does not exist a maximally coherent Gaussian pure state $|\psi\rangle$ such that $|\psi\rangle\xrightarrow {\text IGO} \sigma$ for any Gaussian state $\sigma$. Here we identify a
 maximally coherent state as a state that
allows for the deterministic generation of all other
Gaussian states by means of incoherent Gaussian
operations. Note that  maximally coherent states are independent of
a specific coherence measure.

Above consequences demonstrate significant differences between discrete-variable systems and Gaussian systems. One key reason for these differences is the fact that Gaussian states are completely specified by their first and second moments. Intuitively,  since determinant of the second moment for any pure Gaussian state is 1, conversion of pure Gaussian states by IGO can be realized by Gaussian unitary operations.


Theorem 3.1 is also a nice tool for conversion of pure Gaussian states because the first and second moments of pure Gaussian states have clear analytic formulas.
In the following, we exhibit conditions for  realizing  conversion of pure states  under {\text IGOs}.

 For $\alpha-$state  $$|\alpha\rangle=e^{-\frac{1}{2}|\alpha|^2}\sum_{n=0}^{+\infty}\frac{\alpha^n}{\sqrt n!}|n\rangle$$ $(\alpha\in {\mathbb C})$,
the most important Gaussian states which are generated by the vacuum state $|0\rangle$ and Weyl displacement operator
${\widehat D}(\alpha)=e^{\alpha{\widehat a}^\dag-{\overline\alpha}{\widehat a}}$,
 $|\alpha\rangle={\widehat D}(\alpha)|0\rangle$. Note that $$d=2({ Re}(\alpha), {\mbox Im}(\alpha))^t,\quad V=I \text{\quad\cite{Weedx}.}$$
Obviously $|\alpha\rangle \notin{\mathcal I}$, i.e., $|\alpha\rangle$ is coherent. By Theorem 3.1, 
one can see that $$|\alpha\rangle\xrightarrow{\text IGO}|\beta\rangle\Leftrightarrow |\alpha|=|\beta|.$$

The most general pure Gaussian state $|\psi\rangle$ of one-mode is a displaced squeezed state obtained by the combined action of Weyl displacement operator
${\widehat D}(\alpha)$ and the squeezing operator $${\widehat S}(\beta)=e^{\frac{1}{2}[\beta{{\widehat a}{^\dag}}^2-{\overline\beta}{\widehat a}^2]},\quad \beta\in{\mathbb C},$$
on the vacuum state $|0\rangle$ \cite{Weedx}: $$|\psi_{\alpha, \beta}\rangle={\widehat D}(\alpha){\widehat S}(\beta)|0\rangle.$$
The first and second moment of $|\psi_{\alpha, \beta}\rangle$ are \cite{Olivx} $$ 2({\text Re}(\alpha), {\text Im}(\alpha)),$$

$$\left(\begin{array}{cc}
                                                                           {\text ch}(2|\beta|)+\cos\theta\ {\text sh}(2|\beta|)& \sin\theta\ {\text sh}(2|\beta|)\\
                                                                            \sin\theta\ {\text sh}(2|\beta|) & {\text ch}(2|\beta|)-\cos\theta\ {\text sh}(2|\beta)\end{array}\right),$$  where $\beta=|\beta|e^{i\theta}, {\text ch}(x)=\frac{e^x+e^{-x}}{2},\ {\text sh}(x)=\frac{e^x-e^{-x}}{2}$ are hyperbolic functions. By Theorem 3.1, one can deal with conversion of pure Gaussian states efficiently.

\vspace{0.1in} A particularly key conversion of coherent states in discrete-variable systems is purification which is the process that extracts pure
coherent states from general states by IOs  \cite{Win, FL1x, Wangx, Regx, Torx}. The importance of purification lies in that the quantum systems are rather susceptible
to imperfect operations such as decoherence \cite{NCX, Prex} which may jeopardize the reliability of quantum coherence and so one key question is to extract coherent states
with high quality for application. Especially, in \cite{FL1x}, Fang and Liu show that it is impossible to exactly transform a full rank
coherent mixed state to a pure output coherent state by IOs, even probabilistically. This builds no-go theorem for coherent mixed states with full rank. An interesting question is how about purification of continuous-variable systems?

\vspace{0.1in}{\bf Theorem 3.2.} {\it For a coherent pure Gaussian state $\sigma(V_2,d_2)$, \hspace{0.1in} if there exist an {\text IGO} $\Phi$ and a Gaussian state $\rho(V_1,d_1)$ such that
$\Phi(\rho(V_1,d_1))=\sigma(V_2,d_2)$, then $\rho(V_1,d_1)$ is a pure state.}

Theorem 3.2 is a parallel result of no-go theorem of purification for coherent mixed states of discrete-variable systems \cite{FL1x}.
It shows strong limit on the efficiency of perfect coherent purification of Gaussian states.

In addition, by Theorem 3.2, there does not exist a maximally coherent mixed Gaussian state which can generate all Gaussian states. Combining this with Theorem 3.1, there is not a maximally coherent Gaussian state which can generate all other Gaussian states by means of IGOs.
It is a peculiarity of Gaussian continuous-variable systems.

\vspace{0.1in}

In practical applications such as evolution of quantum coherence of optical cat states, people need to deal with the mixed input and output states rather than pure ones \cite{Glancy, Mzhang}. We will provide  structural characterization of conversion for mixed Gaussian states in the following. It is an answer to the question of characterizing mixed coherent state manipulation in infinite dimensional systems \cite{DBG}. That is, given two mixed Gaussian states $\rho(V_1,d_1)$ and $\sigma(V_2,d_2)$, when $\rho(V_1,d_1)\xrightarrow {\text IGO}\sigma(V_2,d_2)$ holds true.
 Firstly, we need to classify IGOs  for a clear presentation.
By the definition of IGOs, one can see
any IGO of one-mode  has two kinds of types:

$$T=tO_1,\quad N=\omega I\quad(\text{Type I})$$ with
$ \omega\geq|1-t^2|, \quad\det O_1=1$ and $O_1$ is a $2\times 2$ real orthogonal matrix;
$$T=tO_2,\quad N=\omega I\quad(\text{Type II})$$ with
$ \omega\geq1+t^2,\  \det O_2=-1$ and $O_2$ is a $2\times 2$ real orthogonal matrix.
Secondly, for Gaussian states $\rho(V_1,d_1),\quad \sigma(V_2,d_2)$, there are real orthogonal matrices $U$ and $W$ with $\det U=\det W=1$ such that
$$UV_1U^t=\left(\begin{array}{cc}
           \lambda_1 & 0\\
            0 & \lambda_2\end{array}\right),\  WV_2W^t=\left(\begin{array}{cc}
           \mu_1 & 0\\
            0 & \mu_2\end{array}\right).$$  We also assume $\lambda_1\neq\lambda_2, \|d_1\|\neq 0$ for generality, here $\|d_1\|$ is the Euclidean norm of $d_1$.

Now, we are ready to give our results on conversion of mixed Gaussian states.

\vspace{0.1in}{\bf Theorem 3.3.} {\it For Gaussian states $\rho(V_1,d_1),\quad \sigma(V_2,d_2)$,   $ \hspace{0.1in} \rho(V_1,d_1)\xrightarrow {\text IGO}\sigma(V_2,d_2)$ by   type I IGO if and only if one of the followings holds true  $$\begin{array}{ll}
                                            (i) & V_2=\mu I\quad(\mu\geq 1), \quad d_2=0;\\
                                            (ii) & \left\{ \begin{array}{ccc}
                                                       \frac{\|d_2\|^2}{\|d_1\|^2} & = & \frac{\mu_1-\mu_2}{\lambda_1-\lambda_2},\\
                                                         \frac{\|d_2\|^2}{\|d_1\|^2} & \leq &  \min\{\frac{\mu_1}{\lambda_1}, \frac{1+\mu_1}{1+\lambda_1}\},\\
                                                          1-\mu_1 & \leq & (1-\lambda_1)\frac{\|d_2\|^2}{\|d_1\|^2};\end{array}\right.\\
                                             (iii) & \left\{ \begin{array}{ccc}
                                                       \frac{\|d_2\|^2}{\|d_1\|^2} & = & \frac{\mu_1-\mu_2}{\lambda_2-\lambda_1},\\
                                                         \frac{\|d_2\|^2}{\|d_1\|^2} & \leq &  \min\{\frac{\mu_1}{\lambda_2}, \frac{1+\mu_1}{1+\lambda_2}\},\\
                                                          1-\mu_1 & \leq & (1-\lambda_2)\frac{\|d_2\|^2}{\|d_1\|^2}.\end{array}\right.
                                                   \end{array}$$}

\vspace{0.1in}{\bf Theorem 3.4} {\it For Gaussian states $\rho(V_1,d_1),\quad \sigma(V_2,d_2)$,   $ \hspace{0.1in} \rho(V_1,d_1)\xrightarrow {\text IGO} \sigma(V_2,d_2)$ by   type \text {II} IGO if and only if one of the followings holds true  $$\begin{array}{ll}
                                                 (i) & V_2=\mu I\quad(\mu\geq 1), \quad d_2=0;\\
                                                 (ii) &\left\{ \begin{array}{ccc}
                                                       \frac{\|d_2\|^2}{\|d_1\|^2} & = & \frac{\mu_1-\mu_2}{\lambda_1-\lambda_2},\\
                                                         \frac{\|d_2\|^2}{\|d_1\|^2} & \leq &   \frac{\mu_1-1}{\lambda_1+1};\\
                                                          \end{array}\right.\\
                                                    (iii) & \left\{ \begin{array}{ccc}
                                                       \frac{\|d_2\|^2}{\|d_1\|^2} & = & \frac{\mu_1-\mu_2}{\lambda_2-\lambda_1},\\
                                                         \frac{\|d_2\|^2}{\|d_1\|^2} & \leq &  \frac{\mu_1-1}{\lambda_2+1}.\\
                                                          \end{array}\right.
                                                        \end{array}$$}

Theorem 3.3 and Theorem 3.4 are very helpful to fulfil conversion of Gaussian states under IGOs. By \cite{Weedx}, the most general one-mode Gaussian state has the second moment
$$V=(2{\overline n}+1)R(\theta)S(2r)R(\theta)^t,$$ $S(2r)=\left(\begin{array}{cc}
                                                    e^{-2r} & 0\\
                                                   0 & e^{2r}\end{array}\right)$, $r\in{\mathbb R}$ is called the squeezing parameter.
Note that $ R(\theta)$ is real orthogonal, for Gaussian states $\rho(V_1,d_1), \sigma(V_2, d_2)$, one can decide whether $\rho(V_1,d_1)$ can be converted into $\sigma(V_2, d_2)$ or vice versa by parameters ${\overline n_i}, r_i , \|d\|_i, i=1, 2$.

\vspace{0.1in}

{\it \bf IV. Conversion of pure Gaussian states of two-mode continuous-variable systems}
\vspace{0.1in}

Originating from Theorem 3.1, an interesting question is to describe conversion of pure Gaussian states of $m$-mode continuous-variable systems $(m\geq 2)$ under IGOs.  We will firstly attempt to discuss two-mode case. The exploratory study reveals a big task for conversion of pure Gaussian states of $m$-mode continuous-variable systems $(m\geq 2).$ The difficulty lies in computational complexity of finding determinant relationship between second moments of  Gaussian states and its blocks. In this section, we discuss the structure of IGOs of two-mode continuous-variable systems further. Based on this, conversion of one kind of important pure Gaussian states under IGOs is described.

For generality, we assume that the second moment of output state is not diagonal.  By the definition of IGOs, it is easy to check that
 IGOs of $2$-mode  have two kinds of types:

$$T=\left(\begin{array}{cc}
        t_1 O_1 & 0\\
       0 & t_{2}O_2\end{array}\right),\quad N=\left(\begin{array}{cc}
                                            \omega_1 I & 0\\
                                            0 & \omega_{2}I\end{array}\right)\quad(\text{Type I})$$ with
$$ \omega_1\geq|1-t_1^2\det O_1|,\quad \omega_2\geq|1-t_2^2\det O_2|; $$

$$T=\left(\begin{array}{cc}
        0 &  t_{2}O_2\\
       t_1 O_1 & 0\end{array}\right), \quad N=\left(\begin{array}{cc}
                                            \omega_1 I & 0\\
                                            0 & \omega_{2}I\end{array}\right) \quad(\text{Type II})$$ with  $$\omega_1\geq|1-t_2^2\det O_2|, \quad\omega_2\geq|1-t_1^2\det O_1|,$$ where $O_1,O_2$ are $2\times 2$ real orthgonal matrices with $\det O_i=\pm 1$, (i=1,2).  We find that if the above IGOs transform one pure state into the other pure state, then $\det O_i=1 (i=1,2)$ as following.

\vspace{0.1in}{\bf Theorem 4.1.} {\it For pure Gaussian states $\rho(V_1,d_1),\quad \sigma(V_2,d_2)$, writing $V_1$ and $V_2$ in their block form
$V_1=\left(\begin{array}{cc}
                V_{11} & V_{12}\\
                V_{12}^t & V_{22}\end{array}\right)$, $V_2=\left(\begin{array}{cc}
                V_{11}' & V_{12}'\\
                V_{12}'^t & V_{22}'\end{array}\right)$, $V_{12}'\neq 0$, where $V_{ij}$ and $V_{ij}'(i,j=1,2)$ are $2\times 2$ matrices, if there exists some $\text{IGO}$ $\Phi$ such that $\Phi(\rho(V_1,d_1))=\sigma(V_2,d_2)$, then $\det O_1=\det O_2=1$.}
\vspace{0.1in}

Theorem 4.1 is useful to realize the conversion of pure Gaussian states. An important class of two-mode Gaussian states has second moments in standard form
$$V=\left(\begin{array}{cc}
                aI & C\\
                C & bI\end{array}\right)\quad {\mbox with} \quad
C=\left(\begin{array}{cc}
                c & 0\\
                0 & d \end{array}\right),$$ $a\geq 1, b\geq 1, c,d\in{\mathbb R}$ \cite{Weedx, Duan1, Simon1}. Any Gaussian state can be transformed to the
Gaussian state with the second moment in standard form by local linear unitary Bogoliubov operations \cite{Duan1}. We will give a complete classification of conversion for such kind of pure states. One can check that such Gaussian states  are pure if and only if $$ab-c^2> 0, (ab-c^2)(ab-d^2)=1, a^2+b^2+2cd\leq 2.$$
Let $$V'=\left(\begin{array}{cc}
                a'I & C'\\
                C' & b'I\end{array}\right), \quad {\mbox with} \quad
C'=\left(\begin{array}{cc}
                c' & 0\\
                0 & d' \end{array}\right)$$  be the second moment of pure target states.
The key step for realizing $\rho(V,d)\xrightarrow {\text IGO} \sigma(V',d')$ is $$V\xrightarrow {\text IGO} V'.$$ We will firstly classify the transformation on second moments.
Based on this classification, conversion of pure Gaussian states can be investigated easily.

In the following, assume  $cd\neq 0, c'd'\neq 0$ and denote
$\alpha=\frac{{c'}^2}{c^2}$ or $\alpha=\frac{{c'}^2}{d^2}$, there are four important closed intervals which are needed to classify transformation between $V$ and $V'$:
$$\left.\begin{array}{l}
(1)\quad  [\frac{\alpha(b+1)}{b'+1},\quad\min\{1,\frac{a'-1}{a-1},\alpha\}],\\
(2)\quad [\max\{\alpha,\frac{\alpha(b-1)}{b'-1}\},\qquad\quad1],\\
(3) \quad[\max\{1,\frac{\alpha(b+1)}{b'+1}\},\quad\min\{\alpha, \frac{a'+1}{a+1}\}],\\
(4)\quad [\max\{\alpha,\frac{\alpha(b-1)}{b'-1}, 1\}, \qquad\frac{a'+1}{a+1}]\end{array}\right.$$

\vspace{0.1in}{\bf Theorem 4.2.} {\it $$V\xrightarrow {\text IGO} V'\Leftrightarrow \Omega\neq\emptyset,$$  here $\Omega=(1)\cup (2)\cup (3)\cup (4)$.}

For $\rho(V,d), \sigma(V',d')$, if $\Omega\neq\emptyset$ and the desired IGO is type I, from the proof of Theorem 4.2 in appendix, then we choose arbitrarily $t_1\in\Omega$. $t_2,\omega_1, \omega_2$ are decided by $t_1$.
One can check easily that whether there exists suitable $T$ such that  $Td=d'$. Therefore the conversion between $\rho(V,d)$ and $\sigma(V',d')$ can be ascertained.
If the IGO is type II, then we pick any $t_2\in\Omega$. The existence of $T$ satisfying $Td=d'$ can be checked directly.


\vspace{0.1in}

\vspace{0.1in}

{\it\bf V. Summary} \vspace{0.1in}

In this work, we have studied conversion of coherent Gaussian states under incoherent Gaussian operations.
An explicit description on conversion of one-mode systems has been provided. Compared with the finite dimensional results on conversion of coherent states \cite {Bau, Chi, Strelt, DBG}, there are some
peculiarities  as following: (1) There does not exist a maximally coherent Gaussian state which can generate all coherent Gaussian states; (2) The conversion between pure Gaussian states is reversible; (3) The coherence of input pure state and the coherence of output pure state are equal. This implies frozen behaviour \cite{Adesoxx} in Gaussian dynamical systems which is listed as an open question in \cite{Xu1}. Conversion of pure Gaussian states of two-mode systems under incoherent Gaussian operations is also discussed. We classify conversion for an important class of two-mode pure Gaussian states with  second moments in the standard form \cite{Weedx, Duan1, Simon1}.

Our results raise some interesting questions. It would be of great interest to classify conversion of pure Gaussian states or mixed Gaussian states for $m$-mode ($m\geq 2$) continuous-variable systems. This is very helpful for comprehending  behaviours of coherence of Gaussian dynamical systems.

\vspace{0.1in}
{\it \bf Acknowledgement}

The authors are indebted to referee for his/her valuable suggestions which improve the presentation of the study.

We acknowledge that the research was  supported by NSF of China (11671332), NSF of
Fujian (2018J01006).

\vspace{0.1in}

{\it\bf Appendix: Proof of main results}\vspace{0.1in}

Proofs of all results in this paper are given in appendix.

The proofs of our theorems need  structural classification of real orthogonal matrices  and determinant formula of sum of two matrices borrowed from
\cite{HJX}.

\textbf{Proposition 1.} {\it Let $R(\theta)=\left(\begin{array}{cc}
                                                 \cos\theta & \sin\theta\\
                                                   -\sin\theta & \cos\theta\end{array}\right)$, where $\theta$ is a real parameter.
$O$ is real orthogonal if and only if $O=R(\theta)$ or
                                                 $O=\left(\begin{array}{cc}
                                                       1 & 0\\
                                                       0 & -1\end{array}\right)R(\theta)$}.

\textbf{Proposition 2.} {\it For $A,B\in{\mathcal M}_2({\mathbb R})$, $$\det(A+B)=\det(A)+\det(B)+tr(A^\sharp B),$$ where $(\cdot)^\sharp$ denotes the adjugate map given by
$$M^\sharp=\left(\begin{array}{cc}
           d & -b\\
           -c & a\end{array}\right),\quad\text{where}\quad M=\left(\begin{array}{cc}
           a & b\\
           c & d\end{array}\right).$$}

The following characterization of Gaussian states can be found in \cite{Serafini1, Serafini2}.

\textbf{Proposition 3.} For any Gaussian state with second moment $V=\left(\begin{array}{cc}
                                                                       A & C\\
                                                                       C^t & B\end{array}\right)$, where $A,B,C$ are real $2\times 2$ matrices. Denoting $\Delta=\det A+\det B+2\det C$, we have
$$V>0,\quad \det V\geq 1,\quad \Delta\leq 1+\det V.$$

\vspace{0.1in}

Now we are in a position to give proofs of our theorems.

{\bf Proof of Theorem 3.1.}
``$\Rightarrow $'' Assume that there exists an IGO $\Phi$ such that $\Phi(\rho(V_1,d_1))=\sigma(V_2, d_2)$. By the definition of IGO, we can obtain
$$t^2 O V_1O^t+\omega I=V_2, \quad tOd_1=d_2, \eqno (1)$$ where $O$ is a real orthogonal matrix, $\omega, t\in {\mathbb R},\  \omega\geq|1-t^2\det O|$.
We divide the proof into two cases by Proposition 1.

Case (i) $O=\left(\begin{array}{cc}
                              1 & 0\\
                              0 & -1\end{array}\right)R(\theta)$.

It is easy to see that $\det O=-1$ and so $\omega\geq 1+t^2$. Combining (1) with Proposition 2, one can get
$$\left.\begin{array}{lll}
           1=\det V_2 & = & t^4\det V_1+\omega^2+\omega t^2 tr(V_1)\\
            & = & t^4+\omega^2+\omega t^2 tr(V_1).\end{array}\right.$$
By $\omega\geq 1+t^2$, we have $t=0, \omega=1$. Therefore $V_2=I, d_2=0$ and so $\sigma(V_2, d_2)=|0\rangle\in {\mathcal I}$.

Case (ii) $O=R(\theta)$ for some $\theta\in {\mathbb R}$.

It is evident that $\det O=1$ and so $\omega\geq |1-t^2|$. Because $\det V_1=1$, we can assume eigenvalues of $V_1$ are $\lambda_1$ and $\frac{1}{\lambda_1} (\lambda_1> 0)$.
By (1) and spectral mapping theorem, we have $$(t^2\lambda_1+\omega)(\frac{t^2}{\lambda_1}+\omega)=1. \quad \eqno (2)$$
It follows that $$\omega^2+\omega(\lambda_1 t^2+\frac{t^2}{\lambda_1})+t^4-1=0.$$
This implies that $$|t|\leq 1.$$  From the relation of root and coefficient of quadratic equation,  we have
$$\omega=\frac{-(t^2 \lambda_1+\frac{t^2}{\lambda_1})+ \sqrt{ \lambda_1^2 t^4+\frac{t^4}{\lambda_1^2} -2t^4+4}}{2}.$$
From $$-(t^2 \lambda_1+\frac{t^2}{\lambda_1})+\sqrt{ \lambda_1^2 t^4+\frac{t^4}{\lambda_1^2} -2t^4+4}\geq 2(1-t^2),$$
it follows that  $$\lambda_1^2 t^4+\frac{t^4}{\lambda_1^2} -2t^4+4\geq [2+t^2(\lambda_1+\frac{1}{\lambda_1}-2)]^2.$$
A direct computation shows that $$t^2(2-\lambda_1-\frac{1}{\lambda_1})\leq (2-\lambda_1-\frac{1}{\lambda_1}).$$
Note that $\lambda_1+\frac{1}{\lambda_1}\geq 2$ and so $$|t|\geq 1.$$ Hence $|t|=1$. By (2), we have $\omega=0$.
From (1), $$V_2 =R(\theta) V_1 R(\theta)^t, d_2=R(\theta) d_1$$ if $t=1$. If $t=-1$, replacing $\theta$ with $\theta+\pi$, we also get the desired.

``$\Leftarrow $'' If $$V_2=I, d_2=0,$$ Taking $T=tR(\theta), t=0, \omega=1,$ then the IGO induced by $T$ and $\omega$ has desired property.
If there exists $R(\theta)$ such that $$V_2 =R(\theta) V_1 R(\theta)^t, d_2=R(\theta) d_1,$$ then, choosing $T=tR(\theta), t=1, \omega=0$, we have the desired.
\vspace{0.1in}

{\bf Proof of Theorem 3.2.}
For coherent Gaussian pure state $\sigma(V_2, d_2)$, suppose that there exists an IGO $\Phi$ with $\Phi(\rho(V_1,d_1))=\sigma(V_2, d_2)$. Therefore
$$t^2 O V_1O^t+\omega I=V_2, \quad tOd_1=d_2, \eqno (3)$$ where $O$ is a real orthogonal matrix, $\omega, t\in {\mathbb R}, \omega\geq|1-t^2\det O|$.
We declare $O=R(\theta)$ for some $\theta\in {\mathbb R}$.  Otherwise
$O=\left(\begin{array}{cc}
                              1 & 0\\
                              0 & -1\end{array}\right)R(\theta)$. This indicates $\det O=-1$ and so $\omega\geq 1+t^2$. By (3) and Proposition 2 , we get
$$1=\det V_2  =  t^4\det V_1+\omega^2+\omega t^2 tr(V_1) \eqno (4).$$
Since $\omega\geq 1+t^2$, we have $t=0, \omega=1$. Therefore $V_2=I, d_2=0$ and so $\sigma(V_2, d_2)=|0\rangle\in {\mathcal I}$, a contradiction.
In (4), for conciseness, denote $tr(V_1)=a, \det V_1=b$. It is evident that $a>1, b\geq 1$. Hence $$\omega^2+at^2\omega+t^4b-1=0.$$ This indicates $|t|\leq 1$ and so $\omega\geq 1-t^2$.
By the relation between root and coefficient of quadratic equation, we also obtain $$ \frac{-t^2 a+\sqrt{t^4a^2-4t^4b+4}}{2}\geq 1-t^2.$$
It is equivalent to $$t^2(1-a+b)\leq 2-a.$$  The proof is divided into three cases in the following.

Case (i) $1-a+b<0$.

In this case, we immediately have $\frac{2-a}{1-a+b}\leq t^2\leq 1$. Thus $1-a+b\leq 2-a$ and so $b\leq 1$. Since $b=\det V_1\geq 1$,  $b=1$ and $\rho(V_1, d_1)$ is pure.

Case (ii) $1-a+b>0$.

This tells $\frac{2-a}{1-a+b}\geq t^2$ and so $a\leq 2$. From $a\geq 2\sqrt b$, it follows that  $b\leq 1$. Therefore  $b=1$ and $\rho(V_1, d_1)$ is pure.

Case (iii) $1-a+b=0$.

By the assumption, we have $0\leq 2-a$. Using the proof of case (ii), we obtain
$\rho(V_1, d_1)$ is pure.

\vspace{0.1in}

{\bf Proof of Theorem 3.3.}
For Gaussian states $\rho(V_1,d_1), \sigma(V_2,d_2)$, let  $U$ and $W$ be real orthogonal matrices with $\det U=\det W=1$ satisfying
$$UV_1U^t=\left(\begin{array}{cc}
           \lambda_1 & 0\\
            0 & \lambda_2\end{array}\right),\  WV_2W^t=\left(\begin{array}{cc}
           \mu_1 & 0\\
            0 & \mu_2\end{array}\right).$$  Note that

 $$ \rho(V_1, d_1)  \xrightarrow {\text IGO}  \sigma(V_2, d_2)$$ if and only if
 $$\rho'(UV_1U^t, Ud_1)      \xrightarrow {\text IGO}       \sigma'(WV_2W^t, Wd_2).$$
Hence without loss of generality,  we may  assume $$V_1=\left(\begin{array}{cc}
           \lambda_1 & 0\\
            0 & \lambda_2\end{array}\right),\  V_2=\left(\begin{array}{cc}
           \mu_1 & 0\\
            0 & \mu_2\end{array}\right).$$

``$\Rightarrow $''
Suppose there exists an IGO $\Phi$ of type I such that $\Phi(\rho(V_1,d_1)=\sigma(V_2,d_2)$, then $$tO d_1=d_2, \quad t^2 OV_1 O^t+\omega I =V_2 ,$$ here $O$ is a real orthogonal matrix with $\det O=1$, $\omega\geq|1-t^2|$ for some real number $t$.
This implies that $$\left\{\begin{array}{l}
t^2=\frac{\|d_2\|^2}{\|d_1\|^2}\\
t^2(\lambda_1-\lambda_2)\sin2\theta = 0\\
t^2(\lambda_1\cos^2\theta+\lambda_2\sin^2\theta)+\omega=\mu_1\\
 t^2(\lambda_1\sin^2\theta+\lambda_2\cos^2\theta)+\omega=\mu_2.\end{array}\right.$$ We divide the proof into two cases.

Case (i) $t=0$.

It is evident that $V_2=\mu I, d_2=0, \mu\geq 1$.

Case (ii) $t\neq 0, \sin2\theta=0$.

We can assume that $\theta=0$ or $\theta=\frac{\pi}{2}$.
If $\theta=0$, then $$\left\{\begin{array}{ccc}
                         t^2\lambda_1+\omega & = & \mu_1\\
                         t^2\lambda_2+\omega & = & \mu_2.\end{array}\right.$$
Hence $$\frac{\mu_1-\mu_2}{\lambda_1-\lambda_2}=\frac{\|d_2\|^2}{\|d_1\|^2},\quad \omega=\mu_1-\frac{\|d_2\|^2}{\|d_1\|^2}\lambda_1.$$
From $\omega\geq 0$, we have $$\frac{\|d_2\|^2}{\|d_1\|^2}\leq\frac{\mu_1}{\lambda_1}.$$
Using $\omega\geq |1-t^2|$, we can obtain $$\frac{\|d_2\|^2}{\|d_1\|^2}\leq\frac{1+\mu_1}{1+\lambda_1},\  1-\mu_1\leq(1-\lambda_1)\frac{\|d_2\|^2}{\|d_1\|^2}.$$
Therefore $$\left\{ \begin{array}{ccc}
                                                       \frac{\|d_2\|^2}{\|d_1\|^2} & = & \frac{\mu_1-\mu_2}{\lambda_1-\lambda_2}\\
                                                         \frac{\|d_2\|^2}{\|d_1\|^2} & \leq &  \min\{\frac{\mu_1}{\lambda_1}, \frac{1+\mu_1}{1+\lambda_1}\}\\
                                                          1-\mu_1 & \leq & (1-\lambda_1)\frac{\|d_2\|^2}{\|d_1\|^2}.\end{array}\right. $$
If $\theta=\frac{\pi}{2}$, then $$\left\{\begin{array}{ccc}
                         t^2\lambda_2+\omega & = & \mu_1\\
                         t^2\lambda_1+\omega & = & \mu_2.\end{array}\right.$$ Hence $$t^2=\frac{\mu_1-\mu_2}{\lambda_2-\lambda_1}=\frac{\|d_2\|^2}{\|d_1\|^2}.$$
Combining $\omega\geq |1-t^2|$ with $\omega=\mu_1-\frac{\|d_2\|^2}{\|d_1\|^2}\lambda_2$, we can get
$$\left\{\begin{array}{l}
      \frac{\|d_2\|^2}{\|d_1\|^2}\leq \frac{\mu_1}{\lambda_2}\\
       \frac{\|d_2\|^2}{\|d_1\|^2}\leq\frac{1+\mu_1}{1+\lambda_2}\\
      1-\mu_1\leq (1-\lambda_2) \frac{\|d_2\|^2}{\|d_1\|^2}.\end{array}\right.$$
Therefore $$\left\{ \begin{array}{ccc}
                                                       \frac{\|d_2\|^2}{\|d_1\|^2} & = & \frac{\mu_1-\mu_2}{\lambda_2-\lambda_1}\\
                                                         \frac{\|d_2\|^2}{\|d_1\|^2} & \leq &  \min\{\frac{\mu_1}{\lambda_2}, \frac{1+\mu_1}{1+\lambda_2}\}\\
                                                          1-\mu_1 & \leq & (1-\lambda_2)\frac{\|d_2\|^2}{\|d_1\|^2}.\end{array}\right.$$

``$\Leftarrow $'' If $V_2=\mu I, d_2=0, \mu\geq 1$, then $t=0, \omega=\mu $ can induce the desired IGO. If
$$\left\{ \begin{array}{ccc}
                                                       \frac{\|d_2\|^2}{\|d_1\|^2} & = & \frac{\mu_1-\mu_2}{\lambda_1-\lambda_2}\\
                                                         \frac{\|d_2\|^2}{\|d_1\|^2} & \leq &  \min\{\frac{\mu_1}{\lambda_1}, \frac{1+\mu_1}{1+\lambda_1}\}\\
                                                          1-\mu_1 & \leq & (1-\lambda_1)\frac{\|d_2\|^2}{\|d_1\|^2},\end{array}\right. $$
then $t^2=\frac{\|d_2\|^2}{\|d_1\|^2}, O=I, \omega=\mu_1-\lambda_1\frac{\|d_2\|^2}{\|d_1\|^2}$ can induce the desired.
If $$\left\{ \begin{array}{ccc}
                                                       \frac{\|d_2\|^2}{\|d_1\|^2} & = & \frac{\mu_1-\mu_2}{\lambda_2-\lambda_1}\\
                                                         \frac{\|d_2\|^2}{\|d_1\|^2} & \leq &  \min\{\frac{\mu_1}{\lambda_2}, \frac{1+\mu_1}{1+\lambda_2}\}\\
                                                          1-\mu_1 & \leq & (1-\lambda_2)\frac{\|d_2\|^2}{\|d_1\|^2},\end{array}\right.$$
then one can choose $t^2=\frac{\|d_2\|^2}{\|d_1\|^2}, O=\left(\begin{array}{cc}
                                               0 & -1\\
                                               1  & 0\end{array}\right), \omega=\mu_1-\lambda_2\frac{\|d_2\|^2}{\|d_1\|^2}$.

{\bf Proof of Theorem 3.4.} Using the same arguments as the start of  proof of Theorem 3.3, we suppose  $$V_1=\left(\begin{array}{cc}
                                                                                    \lambda_1 & 0\\
                                                                                     0 & \lambda_2\end{array}\right), V_2=\left(\begin{array}{cc}
                                                                                    \mu_1 & 0\\
                                                                                     0 & \mu_2\end{array}\right).$$

``$\Rightarrow $'' Assume there exists an IGO $\Phi$ of type II such that $\Phi(\rho(V_1,d_1)=\sigma(V_2,d_2)$, then $$tO d_1=d_2, \quad t^2 OV_1 O^t+\omega I =V_2 ,$$ here $O$ is a real orthogonal matrix with $\det O=-1$, $\omega\geq1+t^2$ for some real number $t$.
This implies that $$\left\{\begin{array}{l}
t^2=\frac{\|d_2\|^2}{\|d_1\|^2}\\
t^2(\lambda_1-\lambda_2)\sin2\theta = 0\\
t^2(\lambda_1\cos^2\theta+\lambda_2\sin^2\theta)+\omega=\mu_1\\
 t^2(\lambda_1\sin^2\theta+\lambda_2\cos^2\theta)+\omega=\mu_2.\end{array}\right.$$ We divide the proof into two cases.

Case (i) $t=0$.

It is evident that $V_2=\mu I, d_2=0, \mu\geq 1$.

Case (ii) $t\neq 0, \sin2\theta=0$.

We assume $\theta=0$ or $\theta=\frac{\pi}{2}$ and will treat them separately.
If $\theta=0$,
then $$\left\{\begin{array}{ccc}
                         t^2\lambda_1+\omega & = & \mu_1\\
                         t^2\lambda_2+\omega & = & \mu_2.\end{array}\right.$$
Therefore $$\frac{\mu_1-\mu_2}{\lambda_1-\lambda_2}=\frac{\|d_2\|^2}{\|d_1\|^2},\quad \omega=\mu_1-\frac{\|d_2\|^2}{\|d_1\|^2}\lambda_1.$$
From $\omega\geq 1+t^2$, we have $\frac{\|d_2\|^2}{\|d_1\|^2}\leq \frac{\mu_1-1}{\lambda_1+1}$, as desired.
If $\theta=\frac{\pi}{2}$, then $$\left\{\begin{array}{ccc}
                         t^2\lambda_2+\omega & = & \mu_1\\
                         t^2\lambda_1+\omega & = & \mu_2.\end{array}\right.$$ Hence $$t^2=\frac{\mu_1-\mu_2}{\lambda_2-\lambda_1}=\frac{\|d_2\|^2}{\|d_1\|^2}.$$
From $\omega=\mu_1-\lambda_2\frac{\|d_2\|^2}{\|d_1\|^2}$ and $\omega\geq 1+t^2$, we obtain $$\frac{\|d_2\|^2}{\|d_1\|^2}\leq \frac{\mu_1-1}{\lambda_2+1}.$$

``$\Leftarrow $'' If $V_2=\mu I, d_2=0, \mu\geq 1$, then $t=0, \omega=\mu $ can induce desired IGO.
If $$\left\{ \begin{array}{ccc}
                                                       \frac{\|d_2\|^2}{\|d_1\|^2} & = & \frac{\mu_1-\mu_2}{\lambda_1-\lambda_2}\\
                                                         \frac{\|d_2\|^2}{\|d_1\|^2} & \leq &   \frac{\mu_1-1}{\lambda_1+1}\\
                                                          \end{array}\right. $$ hold true,
 then $t^2=\frac{\|d_2\|^2}{\|d_1\|^2}, O=\left(\begin{array}{cc}
                                               1 & 0\\
                                               0 & -1\end{array}\right)
, \omega=\mu_1-\lambda_1\frac{\|d_2\|^2}{\|d_1\|^2}$ can induce the desired.
If
$$\left\{ \begin{array}{ccc}
                                                       \frac{\|d_2\|^2}{\|d_1\|^2} & = & \frac{\mu_1-\mu_2}{\lambda_2-\lambda_1}\\
                                                         \frac{\|d_2\|^2}{\|d_1\|^2} & \leq &  \frac{\mu_1-1}{\lambda_2+1}\\
                                                          \end{array}\right.$$  are true, then we take
$t^2=\frac{\|d_2\|^2}{\|d_1\|^2}, O=\left(\begin{array}{cc}
                                               0 & -1\\
                                               -1 & 0\end{array}\right)
, \omega=\mu_1-\lambda_2\frac{\|d_2\|^2}{\|d_1\|^2}$.

{\bf Proof of Theorem 4.1.} Without loss of generality, we assume that $\Phi$ is type I,  the type II case can be treated similarly. By a direct calculation one can obtain $$\begin{array}{ll}
V_2 & =TV_1T^t+N\\
&=\left(\begin{array}{cc}
t_1^2O_1V_{11}O_1^t+\omega_1 I&  t_1 t_2 O_1V_{12}O_2^t\\
  t_1 t_2 O_2 V_{12}^t O_1^t& t_2^2O_2V_{22}O_2^t+\omega_2I\end{array}\right)
\end{array}$$
 Combining Proposition 2 with Proposition 3, we have$$\det( t_1^2O_1V_{11}O_1^t+\omega_1 I)\geq 1,$$ $$\det(t_2^2O_2V_{22}O_2^t+\omega_2I)\geq 1,$$
$$\det(t_1 t_2 O_1V_{12}O_2^t)\leq 0,$$
$$\left.\begin{array}{lll}
\Delta & = & t_1^4\det V_{11}+\omega_1^2+\omega_1t_1^2tr(V_{11})\\
& &+t_2^4\det V_{22}+\omega_2^2+\omega_2t_2^2tr(V_{22})\\
& & +2t_1^2t_2^2\det O_1\det O_2\det V_{12}\\
&\leq  & 2.\end{array}\right.$$
In the following, we divide the proof into two cases .

Case (i) $\det V_{12}'\neq 0$.

Note that $\det V_{12}<0$,  from $$\det(t_1 t_2 O_1V_{12}O_2^t) =t_1^2t_2^2\det V_{12}\det O_1\det O_2\leq 0,$$
we can get $\det O_1\det O_2=1$.  Suppose $\det O_1=-1, \det O_2=-1$, then $\omega_1\geq 1+t_1^2,
\omega_2\geq 1+t_2^2$. This deduces
$$\left.\begin{array}{lll}
\Delta & \geq & t_1^4\det V_{11}+(1+t_1^2)^2+(1+t_1^2)t_1^2tr(V_{11})\\
& &+t_2^4\det V_{22}+(1+t_2^2)^2+(1+t_2^2)t_2^2tr(V_{22})\\
& & + 2t_1^2t_2^2\det V_{12}.\\
&\geq& 2t_1^2t_2^2(\sqrt{\det V_{11}\det V_{12}}+\det V_{12})+2\\
& &+2t_1^2+t_1^4+(1+t_1^2)t_1^2tr(V_{11})\\
& &+ 2t_2^2+t_2^4+(1+t_2^2)t_2^2tr(V_{22}).\end{array}\right.$$
Since $V>0$, we know that $$\sqrt{\det V_{11}\det V_{22}}+\det V_{12}\geq 0.$$ This implies that $\Delta=2$ and so $t_1=t_2=0$, a contradiction.
Hence $\det O_1=\det O_2=1$, as desired.

Case (ii) $\det V_{12}'= 0$.

By Proposition 3,   $$1=\det( t_1^2O_1V_{11}O_1^t+\omega_1 I)=\det(t_2^2O_2V_{22}O_2^t+\omega_2I).$$
Thus $$t_1^4\det V_{11}+\omega_1^2+\omega_1t_1^2tr(V_{11})=1.$$
This tells $|1-t_1^2\det O_1|\leq \omega_1\leq 1$. So $\det O_1=1$, otherwise $t_1=0$ and hence $V_{12}'=0$, a contradiction.
Analogously, one can obtain $\det O_2=1$.

\vspace{0.1in}
{\bf Proof of Theorem 4.2.}
``$\Rightarrow $'' Assume the IGO is Type I and  $$TVT^t+N=V' \qquad(5).$$ By Theorem 3.1 and Proposition 1, we suppose $$O_1=\left(\begin{array}{cc}
                                                                                                                              \cos\theta & -\sin\theta\\
                                                                                                                               \sin\theta & \cos\theta\end{array}\right),\quad O_2=\left(\begin{array}{cc}
                                                                                                                              \cos\phi & -\sin\phi\\                                                                                                                               \sin\phi & \cos\phi\end{array}\right)$$ for some real numbers $\theta,\phi$.
A direct computation of (5) shows that $$\left\{\begin{array}{cc}
                                       c\cos\theta\sin\phi-d\sin\theta\cos\phi & =0\\
                                         c\sin\theta\cos\phi-d\cos\theta\sin\phi & =0.\end{array}\right.$$
Therefore $c^2=d^2$ or $\cos\theta=\cos\phi=0$ or $\sin\theta=\sin\phi=0$. We divide the proof into three cases.

Case (i) $c^2=d^2$

From (5) and Proposition 3, the following equations hold true $$c=-d,$$ $$ c\cos\theta\sin\phi+c\sin\theta\cos\phi=0,$$ $$ ct_1t_2 \cos\theta\cos\phi-ct_1t_2\sin\theta\sin\phi=c',$$ and
$$ ct_1t_2\sin\theta\sin\phi-ct_1t_2\cos\theta\cos\phi=d'.$$
Thus $\sin(\theta+\phi)=0$ and so $\pm ct_1t_2=c'=-d'$. Denoting $\alpha=\frac{{c'}^2}{c^2}$,
comparing the entry of diagonal position of (5), one can obtain
$$\left\{\begin{array}{ll}
         at_1^2+\omega_1=a',& \omega_1\geq |1-t_1^2|\\
         bt_2^2+\omega_2=b', & \omega_2\geq |1-t_2^2|\\
         t_1^2t_2^2=\alpha.\end{array}\right.\quad \eqno (6) $$
If $t_1^2\leq 1, t_1^2\leq\alpha$, then
$$\left\{\begin{array}{l}
               \omega_1=a'-at_1^2\geq 1-t_1^2\\
               \omega_2=b'-b\frac{\alpha}{t_1^2}\geq \frac{\alpha}{t_1^2}-1.\end{array}\right.$$
Hence $$\frac{\alpha(b+1)}{b'+1}\leq t_1^2\leq\min\{1, \frac{a'-1}{a-1}, \alpha\}.$$
If $t_1^2\leq 1, t_1^2\geq\alpha$, then $b't_1^2-b\alpha\geq t_1^2-\alpha.$ Therefore $$\max\{\alpha, \frac{\alpha(b-1)}{b'-1}\}\leq t_1^2\leq \min\{1,\frac{a'-1}{a-1}\}.$$
If $t_1^2>1, t_1^2\leq \alpha$, then $$\left\{\begin{array}{l}
               \omega_1=a'-at_1^2\geq t_1^2-1\\
               \omega_2=b'-b\frac{\alpha}{t_1^2}\geq \frac{\alpha}{t_1^2}-1.\end{array}\right.$$
So $$\max\{1,\frac{\alpha(b+1)}{b'+1}\}\leq t_1^2\leq\min\{\alpha, \frac{a'+1}{a+1}\}.$$
If $t_1^2>1, t_1^2\geq \alpha$, then $b't_1^2-b\alpha\geq t_1^2-\alpha$. Thus
$$\max\{\alpha, \frac{\alpha(b-1)}{b'-1},1\}\leq t_1^2\leq \frac{a'+1}{a+1}.$$

Case (ii) $\sin\theta=\sin\phi=0$.

By a direct computation,  one can get the same intervals and $\alpha$ as case (i).

Case (iii) $\cos\theta=\cos\phi=0$.

In this case, one can get the same intervals as case (i) and $\alpha=\frac{{c'}^2}{d^2}=\frac{{d'}^2}{d^2}$ by similar calculation.
\vspace{0.1in}

If the IGO is type II, a direct computation shows that we have the same intervals and $\alpha$ as Type I. The only difference is that $t_1$ is replaced  by $t_2$.

``$\Leftarrow $'' If the desired IGO is type I, then one can choose $t_1\in \Omega$.  $t_2, \omega_1, \omega_2$ are fixed by (6). Next, according to the interval that $t_1$ belongs, we pick suitable $\theta$ and $\phi$ to
construct some IGO for conversion of pure states.  If the desired IGO is type II, then one can choose $t_2\in \Omega$ and other parameters can be chosen analogously.
$\vspace{0.1in}$


\begin{thebibliography}{99}


\bibitem{3Hor} R. Horodecki, P. Horodecki, M. Horodecki, K.Horodecki,  Rev.  Mod. Phys. \textbf{81}, 865 (2009).

\bibitem{Benett} C.H. Bennett, G. Brassard, C. Crepeau, R. Jozsa, A. Peres, W.K. Wootters, Phys. Rev. Lett.  \textbf{70}, 1895 (1993).

\bibitem{Ein} A. Einstein, B. Podolsky, N. Rosen, Phys. Rev. \textbf{47}, 777 (1935).

\bibitem{VPRK} V. Vedral, M.B. Plenio, M.A. Rippin, P.L. Knight, Phys. Rev. Lett.  \textbf{78}, 2275 (1997).

\bibitem{Virman} M. B. Plenio, S. Virmani, Quan. Inf. Comp. \textbf{7}, 1 (2007).

\bibitem{Abe} J. \AA berg, arXiv:0612146 (2006).

\bibitem{Bau} T. Baumgratz, M. Cramer, M. B. Plenio,
Phys. Rev. Lett.  \textbf{113},  140401 (2014).

\bibitem{Win} A. Winter, D. Yang,  Phys. Rev. Lett.  \textbf{116}, 120404 (2016).

\bibitem{Strelt} A. Streltsov, G. Adesso, M.B. Plenio,  Rev.  Mod. Phys. \textbf{89}, 041003 (2017).

\bibitem{Ben} A. Streltsov, S. Rana, P. Boes, J. Eisert, Phys. Rev. Lett.  \textbf{119},  140402 (2017).

\bibitem{Lud} L. Lami, B. Regula, G. Adesso, Phys. Rev. Lett.  \textbf{122},  150402 (2019).

\bibitem{Chi} E. Chitambar,  G. Gour, Rev. Mod. Phys. \textbf{91} 025001 (2019).

\bibitem{DBG} S. Du, Z. Bai, and Y. Guo, Phys. Rev. A. \textbf{91}, 052120 (2015).

\bibitem {GT1} G. Torun, A. Yildiz, Phys. Rev. A. \textbf{97}, 052331 (2018).

\bibitem {GT2} G. Torun, H.T. Senyasa, A. Yildiz, Phys. Rev. A. \textbf{103}, 032416 (2021).

\bibitem{Gourx} E. Chitambar, G. Gour, Phys. Rev. A \textbf{94}, 052336 (2016).

\bibitem{Benx} A. Streltsov, S. Rana, P. Boes, J. Eisert, Phys. Rev. Lett. \textbf{119},  140402 (2017).

\bibitem{HLS} H. Shi et al., Sci. Rep. \textbf{7}, 14806 (2017).

\bibitem{Du2x} S. Du, Z. Bai, and X. Qi,  Phys. Rev. A \textbf{100}, 032313 (2019).

\bibitem{LZhou1x} C.L. Liu and D.L. Zhou,  Phys. Rev. Lett. \textbf{123}, 070402 (2019).

\bibitem{FL1x} K. Fang and Z. W. Liu, Phys. Rev. Lett. \textbf{125}, 060405 (2020).

\bibitem {Glaux} R.J. Glauber,  Phys. Rev. \textbf{131}, 2766 (1963).

\bibitem{Sudx} E.C.G. Sudarshan, Phys. Rev. Lett. \textbf{10}, 277 (1963)

\bibitem{Braux} S.L. Braunstein and P. van Loock, Rev. Mod. Phys. \textbf{77}, 513 (2005).

\bibitem{Weedx} C. Weedbrook, S. Pirandola, R. Garcia-Patron, N.J. Cerf, T.C. Ralph, J.H. Shapiro, S. Lloyd, Rev. Mod. Phys. \textbf{84}, 621 (2012).

\bibitem{Gie}G. Giedke and J.I. Cirac, Phys. Rev. A 66, 032316 (2002).
\bibitem{Eis3} J. Eisert, S. Scheel and M.B. Plenio, Phys. Rev. Lett. 89,
137903 (2002).
\bibitem{Fiu} J. Fiur\'{a}\v{s}ek, Phys. Rev. Lett. 89, 137904 (2002).
\bibitem{Leo} U. Leonhardt, Measuring the quantum state of light (Cambridge
UP, Cambridge, 1997).

\bibitem{Eis2}J. Eisert and M. B. Plenio, 
Int. J. Quantum. Inform. 01, 479
(2003).


\bibitem {Glancy} S. Glancy, H.M. Vasconcelos, T.C. Ralph, Phys.Rev.A, \textbf{70}, 022317 (2004).

\bibitem  {Mzhang} M. Zhang, H. Kang, M. Wang, F. Xu, X. Su, K. Peng, Photonics Research \textbf{9}, 887 (2021).


\bibitem{Eis1} J. Eisert, C. Simon, and M. B. Plenio, J. Phys. A 35, 3911
(2002).

\bibitem{Xu1} J. Xu, Phys. Rev. A  \textbf{93}, 032111 (2016).

\bibitem {Danx} B. Daniela and G. Nocerino, etc., arXiv:1609.00913 (2016).

\bibitem{Xu2} J. Xu, Phys. Lett. A \textbf{387}, 127028 (2021).


\bibitem{Adesoxx} T.R.Bromley, M.Cianciaruso,and G.Adesso, Phys. Rev. Lett. \textbf{114}, 210401 (2015).

\bibitem{Olivx} S. Olivares, Eur. Phys. J. \textbf{203} ,3 (2012).

\bibitem{Wangx} K. Fang, X. Wang, L. Lami, B. Regula,
and G. Adesso, Phys. Rev. Lett. \textbf{121}, 070404 (2018).

\bibitem{Regx} B. Regula, K. Fang, X. Wang, and G.
Adesso,  Phys. Rev.
Lett. \textbf{121}, 010401 (2018).

\bibitem{Torx} G. Torun, L. Lami, G.Adesso and A.Yildiz, Phys.Rev.A, \textbf{99}, 012321 (2019).

\bibitem{NCX} M. A. Nielsen and I. L. Chuang, Quantum Computation
and Quantum Information: 10th Anniversary Edition,  (Cambridge University Press, New
York, NY, USA, 2011)

\bibitem {Prex} J. Preskill,  Quantum \textbf{2}, 79 (2018).


\bibitem{Duan1} L.M. Duan, G. Giedke, J.I. Cirac and P. Zoller, Phys. Rev.
Lett. \textbf{84}, 2722 (2000).


\bibitem {Simon1} R. Simon, Phys. Rev.
Lett. \textbf{84}, 2726 (2000).

\bibitem{HJX} R. Horn, C. Johnson, Matrix Analysis, (Cambridge University Press, Cambridge, 1986)

\bibitem{Serafini1} A. Serafini, Phys. Rev. Lett. \textbf {96}, 110402 (2006).


\bibitem{Serafini2} S. Pirandola, A. Serafini,and S. Lloyd, Phys. Rev. A \textbf{79}, 052327 (2009).







\end{thebibliography}
\end{document}